\documentclass{aastex631}

\usepackage{float}
\usepackage{xcolor}
\begin{document}

\title{Constraining the Mass Loss and the Kinetic Energy of Solar Coronal Mass Ejections with Far-Ultraviolet Flares}

\author[0000-0002-3153-0157]{Nuri Park}
\affiliation{School of Earth and Space Exploration, Arizona State University \\ Tempe, AZ 85287, USA}

\author[0000-0002-7260-5821]{Evgenya L. Shkolnik}
\affiliation{School of Earth and Space Exploration, Arizona State University \\ Tempe, AZ 85287, USA}
\affiliation{Interplanetary Initiative, Arizona State University \\ Tempe, AZ 85287, USA}

\author[0000-0003-4450-0368]{Joe Llama}
\affiliation{Lowell Observatory \\ Flagstaff, AZ 86001,USA}

\begin{abstract}
Stellar eruptive events, such as flares and coronal mass ejections (CMEs), can affect planetary habitability by disturbing the stability of their atmospheres. For instance, strong stellar flares and CMEs can trigger atmospheric escape and, in extreme cases, may strip away the atmosphere completely. While stellar flares have been observed and explored at a wide range of wavelengths, the physical properties of stellar CMEs remain unconstrained due to the difficulty in observing them. In this context, the Sun provides our only window on the potential characteristics of CMEs on Sun-like stars. A correlation between solar X-ray flare peak flux and the mass of flare-associated solar CMEs has been reported using solar data collected during Solar Cycle 23 \citep{aarnio2011solar}. Here, we build upon that work. We extend the correlation into the far-UV (FUV), where stellar flares are, and will continue to be, routinely detected with existing and future FUV observatories by incorporating data spanning two entire Solar Cycles (23 and 24; 1996\textendash2019). Using three different space missions (CMEs from LASCO/SOHO, X-ray flare events from XRS/GOES, and FUV flares from AIA/SDO), we report a correlation between FUV flare peak flux and energy centered at 1600 {\AA} and mass, kinetic energy, and linear speed of flare-associated CMEs. These empirical relations enable estimates of CME masses and kinetic energies from FUV flares on Sun-like stars. While direct stellar CME detections remain elusive, the correlations derived here are likely applicable to Sun-like stars and provide a working framework for evaluating exoplanet atmospheric erosion.
\end{abstract}


\section{Introduction} \label{sec:intro}

Solar flares and coronal mass ejections (CMEs) are the consequences of solar magnetic field dynamics and magnetic reconnection (e.g., \cite{kahler1992solar,forbes2000solar}). These eruptive events can affect the terrestrial environment by triggering geomagnetic storms and interplanetary disturbances, also known as space weather (e.g., \cite{schwenn2006space}). CMEs are eruptions that release massive amounts of ionized gas into interplanetary space. When the released mass collides with planets, it can disturb the stability of the planetary atmosphere (e.g., geomagnetic storms and enhanced ion escape in the ionosphere) \citep{howard2014space, jakosky2015maven, thampi2021impact}. Other prominent eruptive events resulting from solar magnetic dynamics are solar flares. A solar flare is a sudden increase in solar electromagnetic radiation. Solar flares can trigger photoevaporative and photochemical alteration that can potentially cause a considerable loss and compositional changes in the atmosphere by enhancing electromagnetic radiation in X-ray and ultraviolet (UV) (e.g., \cite{lee2018effects,thiemann2018mars}). These events happen on other stars and may cause similar or more dramatic disturbances to planetary atmospheres, potentially affecting the habitability of exoplanets orbiting those stars \citep{airapetian2020impact}.

Since the first detections of exoplanets, more than 5,800 have been confirmed, including those as small and smaller than the Earth (e.g., \cite{campbell1988search,wolszczan1992planetary,mayor1995jupiter,howell2020grand,hill2023catalog}). Determining the habitability of these planets involves characterizing the physical and chemical environment of the corresponding planetary system, which is highly affected by the intensity and variability of host stellar activity \citep{khodachenko2007coronal,lammer2007coronal,segura2010effect,yamashiki2019impact,atri2020stellar,atri2021stellar}. Among the stellar activity, the physical characteristics of stellar flares, such as energy, duration, and frequency from low-mass stars, have been explored in some detail with observations and are incorporated into various numerical models that quantify their impact on the atmospheres of exoplanets (e.g., \citep{tilley2019modeling,barnes2020vplanet,atri2021stellar,do2022contribution,amaral2025impact}). Despite the fact that we know the Sun produces CMEs at the rate of several per week on average, stellar CMEs are generally not considered in most models due to the lack of conclusive stellar CME detections. 

Although several candidate stellar CMEs have been reported, characterizing their physical properties remains challenging due to current observational limitations \citep{moschou2019stellar,veronig2021indications,wang2021detection,loyd2022constraining}. For instance, many key parameters of solar CMEs—such as mass, velocity, and kinetic energy—are quantified using coronagraph images, as CMEs are significantly dimmer than the solar photosphere \citep{brueckner1995large,domingo1995soho}. However, this method is not yet viable for other stars, obscuring the physical characteristics of stellar CMEs. In contrast, there is a wealth of CME data for the Sun. The Large Angle Spectrometric Coronagraph (LASCO) onboard the Solar and Heliospheric Observatory (SOHO) has observed more than 37,000 CMEs since 1996 \citep{gopalswamy2009soho}. In the absence of direct observations of stellar CMEs, the Sun remains our most detailed and informative laboratory.

One potential approach to estimating the mass of stellar CMEs is by using a flare-CME correlation derived from solar data. It is an ongoing topic of active research on whether flares trigger CMEs or CMEs trigger flares, but it is understood that they are associated with the same magnetic reconnection event. CMEs that occur with solar flares are called flare-associated CMEs. \cite{andrews2003search} found that 58\% of high-intensity X-ray solar flares (X-ray flare peak flux above $10^{-5} \rm W/m^{2}$) that occurred from 1996 to 1999 were associated with CMEs. \cite{aarnio2011solar} derived a linear correlation between the log mass of solar CMEs and log solar X-ray flare peak flux using solar data from 1996 to 2006, constraining the extent of stellar mass release when stellar flares are observed. For instance, the solar X-ray flare-CME correlation was applied to estimate mass loss from young active stars \citep{aarnio2012mass}. This flare-CME correlation is limited to X-ray wavelengths. A complementary wavelength that is more readily observable to detect flares and characterize the particle environment of exoplanetary system is the FUV. Due to its high sensitivity to the stellar upper-atmosphere activity consisting of chromosphere, transition region, and coronal emission, stellar flares from other stars are routinely observed in the FUV (e.g., \cite{loyd2018hazmat,loyd2018muscles,jackman2021stellar}).

In this work, we extend the X-ray flare peak flux and CME mass correlation of \cite{aarnio2011solar} to two complete Solar Cycles (Cycle 23 and 24; 1996-2019), as well as generate additional correlations for CME kinetic energy and linear speed with X-ray and FUV flare peak flux and integrated energy. These correlations can be applied to Sun-like stars with detected X-ray or FUV flares, offering a way to constrain CME environments that would otherwise remain inaccessible. 

\section{Method} \label{sec:method}

We adopt the flare-CME pairing method from \cite{aarnio2011solar}, who used temporal and spatial constraints (CMEs that start between 10 to 80 minutes after a flare start time within $\pm45^{\circ}$ angular separations) to find flare-associated CMEs from the LASCO CME catalog using Geostationary Operational Environmental Satellite (GOES) X-ray Spectrometer (XRS) Flare Report\footnote{\url{https://www.ngdc.noaa.gov/stp/space-weather/solar-data/solar-features/solar-flares/x-rays/goes/xrs/}}. Then, we extracted the FUV peak flux during the CME-associated flare event from the Atmospheric Imaging Assembly (AIA) onboard Solar Dynamics Observatory (SDO) data \citep{lemen2011atmospheric}. 

SDO was launched in February 2010 with three primary scientific instruments (Atmospheric Imaging Assembly (AIA), Extreme-UV (EUV) Variability Experiment (EVE), and Helioseismic and Magnetic Imager (HMI)) to characterize the origin and the structures of solar magnetic fields \citep{pesnell2012solar}. Among the nine bands of AIA, which include bands that are centered at 94 \AA, 131 \AA, 171 \AA, 193 \AA, 211 \AA, 304 \AA, 335 \AA, 1600 \AA, and 1700 \AA, we used the AIA1600 centered at 1600 \AA. The data cadence of 24 seconds provides detailed FUV information on solar X-ray events \citep{lemen2011atmospheric,boerner2012initial}. However, \cite{aarnio2011solar} used solar CMEs and solar X-ray flares that occurred from 1996 to 2006, which do not overlap with the operation time of SDO since it started operating only in May 2010 and continues to observe today. To overcome this limit, we used the same LASCO CME catalog and Solar Flare Report from GOES for an extended period (1996 to 2019) to incorporate solar CME-associated flares that were simultaneously monitored by SDO.

\subsection{The LASCO CME catalog}

The CME mass values reported in the LASCO catalog are calculated using white light images of the CME. The calculation procedures involved 1) determining the total area of CME in the image, 2) converting the pixel values into mass values through modeling of the Thomson scattering, and 3) integrating all the values within the CME area. The kinetic energy values of the LASCO CMEs were calculated by using linear speed and mass values, where linear speed is derived by fitting a linear line to the height-time measurements.

In total, we started with 21,945 CMEs that occurred from 1996 to 2019 in this study, including 708 halo CMEs, which are labeled as such when the CMEs are propagated towards or away from the line of sight of the instrument, see Figure 1.

\begin{figure}[hbt!]
\centering
\includegraphics[width=100mm]{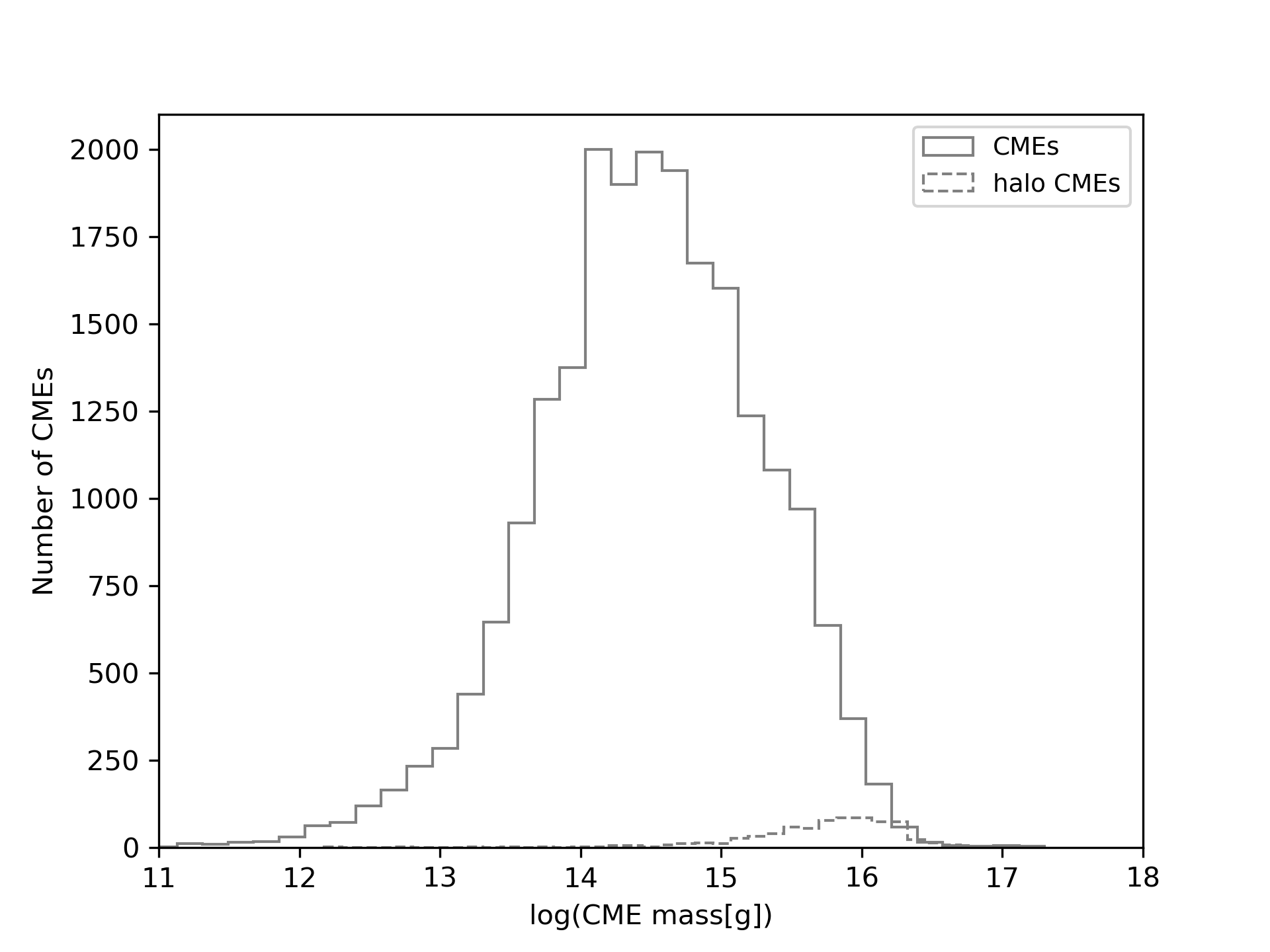}
\caption{The mass distribution of solar CMEs in the Large Angle Spectrometric Coronagraph (LASCO) CME catalog dataset that occurred from 1996 to 2019. The mass values of CMEs that propagated towards or away from the line of sight of the instrument are labeled as halo CMEs in the LASCO CME database. The mass distribution of halo CMEs is presented as a dashed histogram. In total, we started with 21,945 LASCO CMEs, including 708 halo CMEs, in our analyses.  
\label{fig:general}}
\end{figure}

\subsection{Solar X-ray Flares from GOES XRS Flare Report}

Solar flares are determined by X-ray flare peak flux detected by the X-ray Spectrometer (XRS) onboard GOES. XRS observes the disk-integrated flux of solar X-rays. The peak flux of each flare is measured by the XRS-B channel (1-8 \AA). We exclude flares from the database without position values as we require positions of flares to pair them with CMEs. The position values of flares are given in the Stonyhurst coordinate system (Cartesian) in the GOES database. The coordinates of the flare positions were converted to spherical coordinates to compare with CME central position angles \citep{thompson2006coordinate}. In total, we started with 18,678 GOES solar X-ray flares in this study, see Figure 2.  

For the flare energy values, we applied the flare energy calculation method from \cite{aarnio2012mass} for both X-ray and FUV flares, where they used the flare peak flux value and flare duration time from the GOES Flare Report by assuming linear decay of a flare.

\begin{figure}[hbt!]
\centering
\includegraphics[width=100mm]{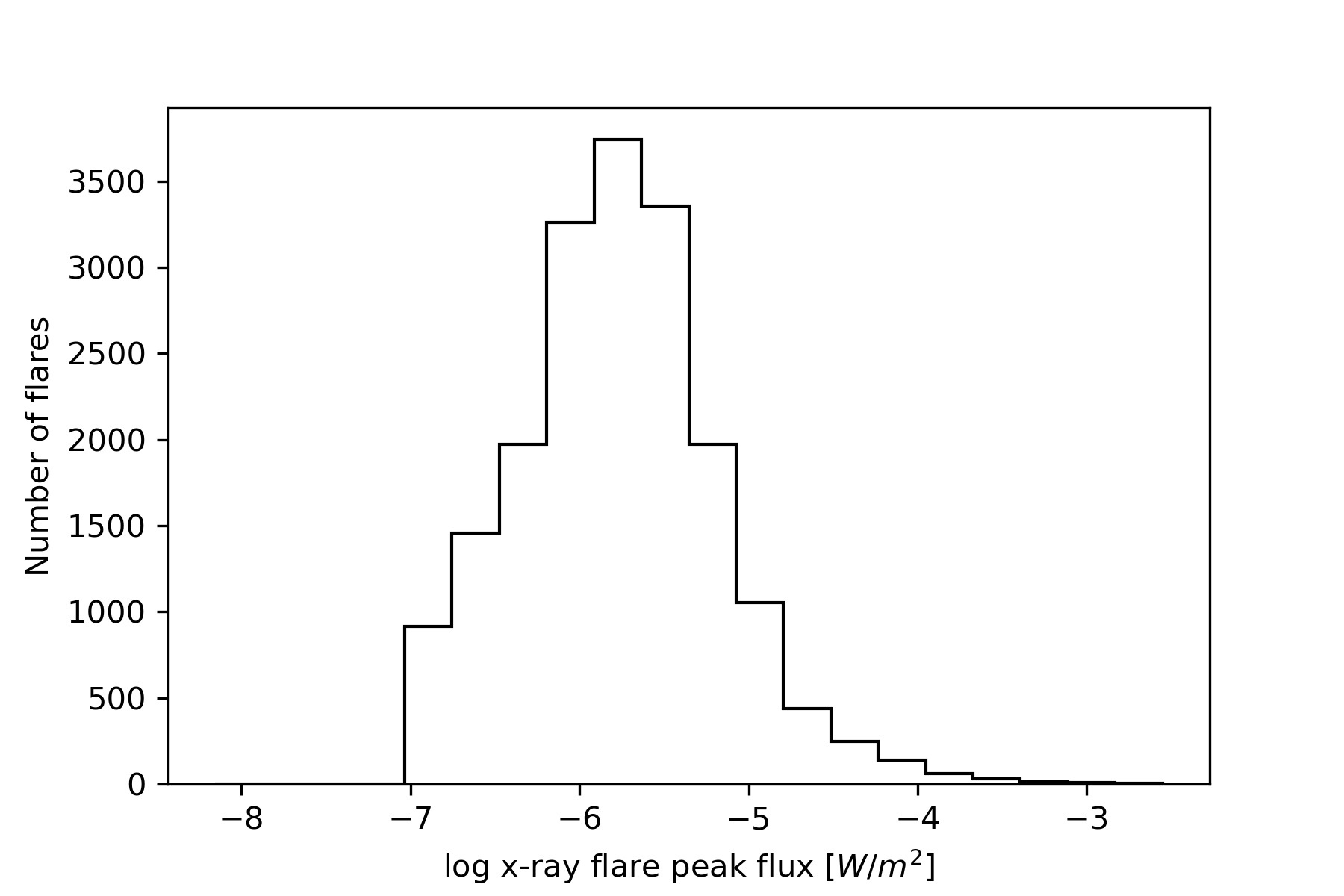}
\caption{The solar X-ray flare peak flux distribution of the GOES solar flare report from 1996 to 2019. We excluded flares without position values, since we require flares that are spatially and temporally coincident with CMEs. 
\label{fig:general}}
\end{figure}

\subsection{Temporal and Spatial Constraints of X-ray Flare-associated CMEs}
\cite{aarnio2011solar} determined the spatial and temporal constraints to find flare-associated CMEs from the LASCO CME catalog by investigating the peak number of CMEs paired within the range of position angle differences and event start time differences. We found a similar peak in the number of CMEs in both constraints, indicating that these constraints cover the range of solar X-ray flare-associated CMEs throughout this extended period (1996-2019), see Figure 3.  

\cite{aarnio2011solar} provided two different linear correlations. First, they excluded halo CMEs when performing flare-CME pairing. Throughout this process, they found that only 6\% of intense X-ray flares (solar flares with flare peak flux higher than $10^{-5} \rm W/m^{2}$) were associated with CMEs. This did not match the result from \cite{andrews2003search}, who found a much higher flare-CME association rate with these intense flares (58\%) using the manual pairing method. \cite{aarnio2011solar} reintroduced the halo CMEs and provided the second correlation, which is suggested as an upper limit to the ejected mass from the flare-associated CMEs since the mass measurement of halo CMEs has higher uncertainty than the regular CME measurements \citep{gopalswamy2009soho}.

The higher uncertainty in halo CME mass measurement is due to the CME mass calculation method, which integrates the pixel values within the manually defined CME shape \citep{vourlidas2000large}. Since the full shape of halo CME appears omnidirectional on the plane of the sky, when the CME is headed towards or away from the instrument, the shape of a halo CME is crudely estimated to retrieve the mass value. This process potentially causes an overestimation of halo CME mass (the mass values of halo CMEs are generally larger than regular CMEs, see Figure 1). However, research as to whether the halo CMEs result from the projection effect or are physically a different class of CME is ongoing \citep{kwon2015halo}. As such, we performed solar flare-CME correlation with and without halo CMEs. Because halo CMEs lack position angles, we applied only temporal constraint when including them. 

\begin{figure}[hbt!]
\centering
\includegraphics[width=180mm]{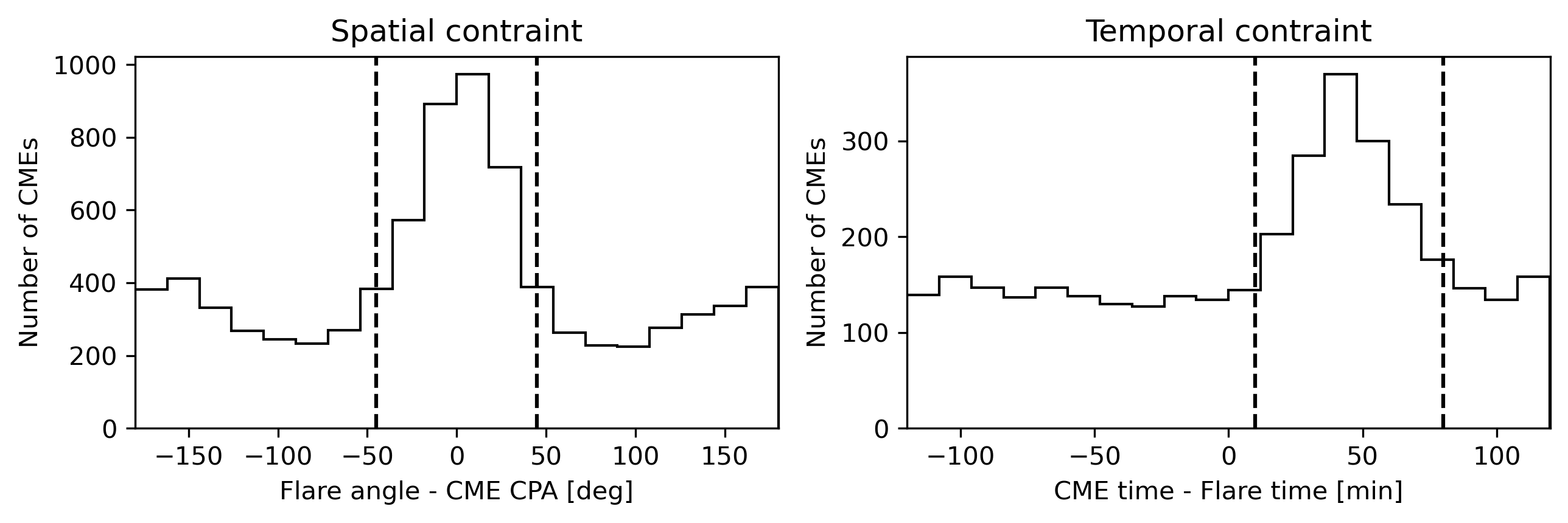}
\caption{(Left) Histogram of solar CMEs occurred within $\pm2$ hours of the solar X-ray flare event start time in the GOES database as a function of angular separation between the corresponding CME and a flare. The dashed lines indicate spatial constraint used to pair flares and CMEs ($\pm45^{\circ}$ of angular separation). (Right) Histogram of selected CMEs within the spatial constraint among the flare-CME pairs that are within $\pm2$ hours start time differences as a function of time differences between the flare event start time and the CME detection time in minutes. The dashed lines indicate temporal constraint; CMEs that started 10 to 80 minutes after the start of a paired X-ray flare event. We found similar distribution peaks observed in \cite{aarnio2011solar}, indicating that these constraints can also be applied to our extended LASCO CME and GOES X-ray flare database. 
\label{fig:general}}
\end{figure}

\subsection{FUV Flux from AIA1600/SDO}
We used AIA1600/SDO images with which to detect FUV flares. We converted AIA1600 images, which cover the C IV line emitted from the transition region of the Sun, to spectral irradiance values using the instrumental information given by \cite{boerner2012initial}. We calculated the number of photons that collided with the AIA1600 CCD, using the following equation: 

\begin{equation}
N_{\rm photon} [{\rm photon}/{\rm m}^{2}] = \frac{p_{\rm total}[{\rm DN}]}{EA(\lambda)[{\rm cm}^{2}] \times {\rm DN}(\lambda) [{\rm DN / electron}] \times QE [{\rm electron/photon}] \times 10^{-4}}
\end{equation}

where $EA(\lambda)$, $DN(\lambda)$, and $QE$ are the effective area, gain in digital number, and quantum efficiency of the AIA1600 instrument, respectively. The calculated number of photons was then incorporated into the energy calculation equation, which corresponds to

\begin{equation}
E[\rm J/m^{2}] = \frac{hc}{\lambda} = \frac{1.986 \times 10^{-16} [\rm J \ nm \ photon^{-1}]}{1600\AA} \times N_{\rm photon}
\end{equation}

The calculated energy density, $E$, was then divided by the exposure time ($t_{\rm exp}$) of the instrument to be converted into FUV flux values ($F$) as 

\begin{equation}
F[\rm W/m^{2}] = \frac{E [\rm J/m^{2}]}{t_{\rm exp} [s]}
\end{equation}

We averaged the converted spectral irradiance values from 24-second to one-minute cadence data to compare them with the event times of the GOES X-ray flare dataset, which is extracted from a one-minute averaged X-ray light curve. The averaged FUV data was then detrended to remove quasi-sinusoidal signals which are the combined result of the Sun's intrinsic rotation and the satellite's motion, see Figure 4. 

\begin{figure}[hbt!]
\centering
\includegraphics[width=180mm]{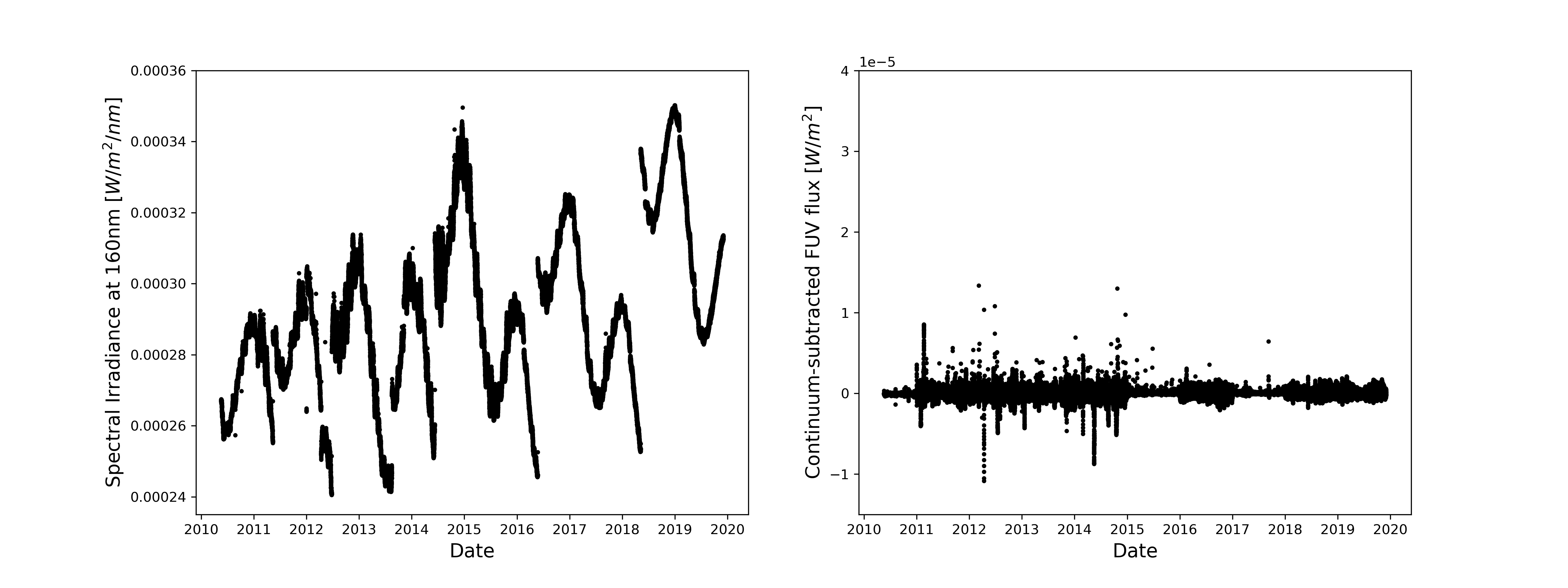}
\caption{One-hour averaged solar spectral irradiance at 1600 \AA (left) and one-hour averaged detrended FUV flux from AIA1600 images (right). For clarity in this 10-year timeseries, one-hour averages are plotted; however, all detrending and quantitative analyses were performed using one-minute averaged data.
\label{fig:general}}
\end{figure}

\section{Results \& Discussion} \label{sec:result}

\subsection{Correlations between X-ray Flare and the X-ray Flare-Associated CMEs}

Among 21,945 LASCO CMEs detected between August of 1996 and December of 2019, we found 2,460 that are associated with X-ray flares ($\approx 11\%$). This number includes halo CMEs. When halo CMEs are excluded from the flare-CME pairing process, we found 2,165 flare-associated CMEs among 21,237 LASCO CMEs ($\approx 10\%$). The flare-CME association rate here is slightly less than that observed by \cite{aarnio2011solar}, who found 1,153 pairs among 7,741 LASCO CMEs ($\approx 15\%$) from 1996 to 2006 when halo CMEs were included. To explore our results in detail, we divided the flare-associated CMEs by Solar Cycle. Since all the flare-CME pairs in our dataset occurred between August 1996 and December 2019, we labeled flare-CME pairs before December 2008 as Cycle 23 flare-CME pairs and the rest as Cycle 24 flare-CME pairs. We found 1,396 and 769 flare-associated CMEs without halo CMEs for Cycle 23 and 24 among 10,998 and 10,239 LASCO CMEs, respectively. This corresponds to $\approx 12.6\%$ and $7.5\%$ of flare-CME association. When halo CMEs are included, we found 1,593 and 867 flare-associated CMEs for Cycle 23 and 24 among 11,394 and 10,551 LASCO CMEs during Cycle 23 ($\approx 13.9\%$) and 24 ($\approx 8.2\%$), respectively.

The Pearson coefficients of the flare-CME pairs indicate weak correlations between CME properties (mass, kinetic energy, and linear speed) and paired X-ray flare peak flux and energy, ranging from 0.2 to 0.4 for both Solar Cycle 23 and Cycle 24, with and without halo CMEs, see Figures 5-8 and Tables 1-4. To reduce the scatter within the dataset, we binned the data along the flare peak flux and flare energy and performed the correlation analyses using the median values of corresponding bins. Specifically, flare peak fluxes (in $W/m^{2}$) were divided into six fixed-width bins spanning 0.5 dex in $log_{10}(flux)$, and flare energies (in erg) were binned into six intervals of 0.7 dex in $log_{10}(energy)$. These fixed log-scale intervals were selected to preserve the physical scaling relationship and allow for consistent comparisons across flare flux and energy regimes. While \cite{aarnio2011solar} and \cite{compagnino2017statistical} employed a binning method that ensured equal sample sizes per bin, our approach emphasizes physical interpretability over uniformity. Although this leads to varying numbers of events per bin, we mitigate statistical bias by using robust measures\textemdash the median and interquartile range (IQR)\textemdash which reduce the influence of outliers and ensure stable central trend estimates. Correlations were then derived using Orthogonal Distance Regression (ODR), allowing for uncertainties in both variables to be incorporated via the IQRs in flare and CME properties. Using this approach, we found linear trends between the median values of flare and CME properties. The derived log-linear correlation equations and the Pearson coefficients for these correlations are presented in Figures 5-8 and Tables 1-4. For applications to other stars, using bins defined by fixed ranges of flare flux or energy may be more appropriate, as this approach preserves the underlying physical scaling and allows for more straightforward comparisons across different stellar environments. In contrast, equal-number binning schemes, while useful for statistical uniformity, may reflect the specific distribution of solar events and could introduce biases when extrapolated to stars with different flare populations.

We note that the regression fits and their $1\sigma$ confidence regions for the flare energy-CME correlation plots (Figures 7 and 8) were obtained using mean-centered flare energy values\textemdash that is, by shifting the x-axis values to be around the mean of the median values of the corresponding bins. This procedure improves the numerical conditioning of the regression, yielding more reliable estimates of the slope, intercept, and associated confidence intervals. The resulting confidence regions are therefore not artificially inflated by the large absolute values of $log_{10}(energy)$ $\sim$ 25-32, as would be the case if the fits were performed without centering. Importantly, this adjustment does not alter the underlying scatter of the data and the fit, but ensures that the reported $1\sigma$ intervals reflect the variability of the flare-CME correlations.

The correlations reported by \cite{aarnio2011solar} for the mass-flare peak flux and by \cite{aarnio2012mass} for the mass-flare energy mostly reside within the IQR of each bin in our analysis, indicating that the difference between these correlations is not significant, see Figure 5 for the CME mass and X-ray flare peak flux correlation and Figure 7 for the CME mass and X-ray flare energy correlation. Considering that solar X-ray flares and CMEs used in \cite{aarnio2011solar} and \cite{aarnio2012mass} represent Solar Cycle 23 (1996-2008) and our data includes the whole Solar Cycle 23 and Cycle 24, our results indicate that these correlations are not highly affected by the Solar Cycle.

However, when halo CMEs are excluded, our correlations predict lower CME mass at higher flare peak flux and flare energy when compared to the correlation from \cite{aarnio2011solar} and \cite{aarnio2012mass}. Considering that these correlations from \cite{aarnio2011solar} and \cite{aarnio2012mass} both include halo CMEs, the halo CMEs presented in Figures 5 and 7 are generally paired with relatively high peak flux and energy X-ray flares. Due to this effect, pairing flares and CMEs without halo CMEs results slightly less values of the Pearson coefficients.

In addition, we found no significant differences in the Pearson coefficients during Solar Cycle 23 and Cycle 24 among CME properties and X-ray flare peak flux/energy dataset. As indicated by comparing the correlations from \cite{aarnio2011solar,aarnio2012mass} and our correlations for flare-CME pairs with halo CMEs, there were no dramatic changes in the Pearson coefficient when comparing Cycle 23 and 24 without halo CMEs, implying that the physical properties of flare-CME pairs remain consistent regardless of the Solar Cycle. 

\begin{figure}[H]
\centering
\includegraphics[width=180mm]{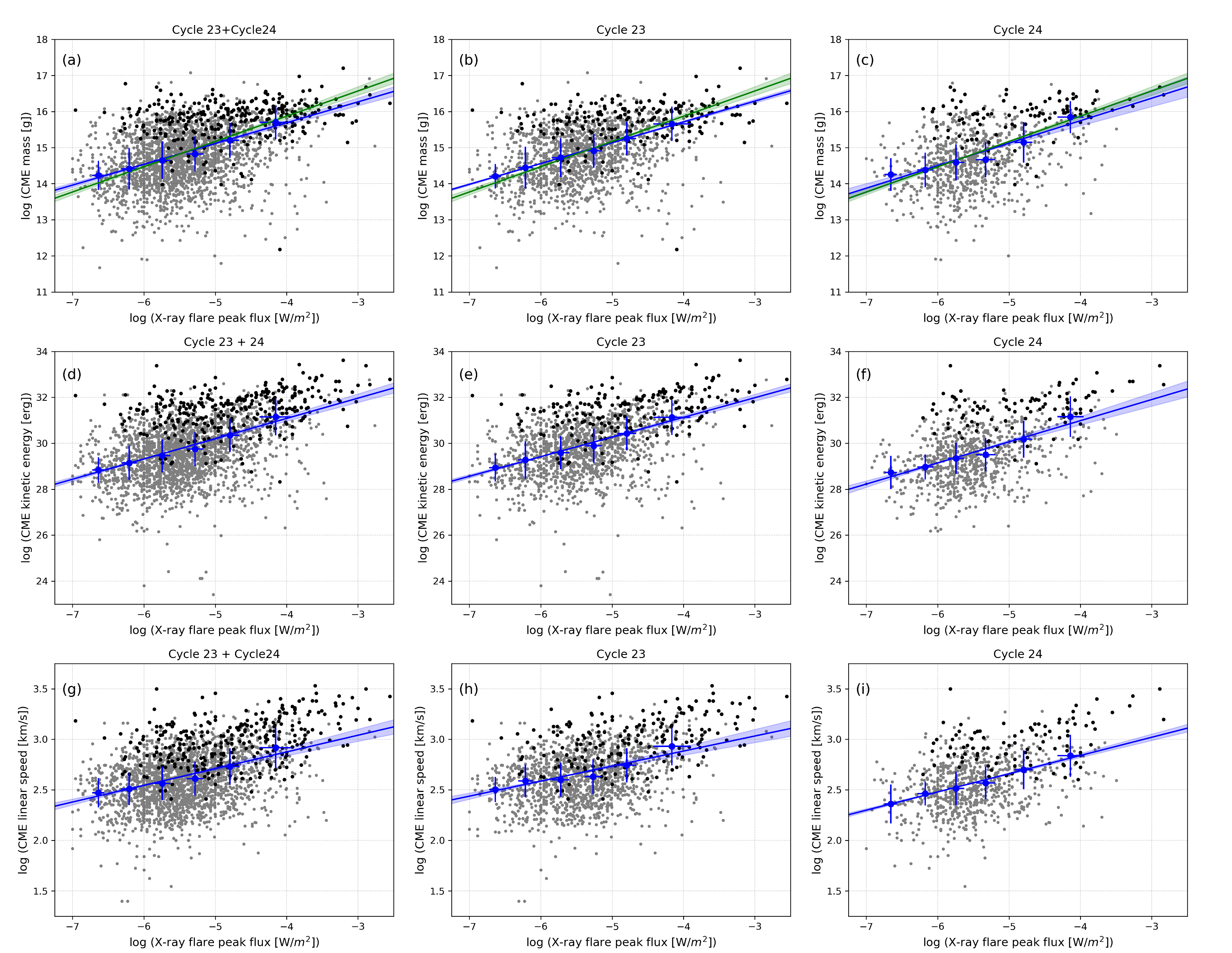}
\caption{The mass of the flare-associated LASCO CMEs with halo CMEs plotted along the X-ray peak flux of associated GOES X-ray flares that occurred during Solar Cycle 23 and 24 (a), Cycle 23 (b), and Cycle 24 (c), the kinetic energy of the corresponding CMEs along the X-ray peak flux during Cycle 23 and 24 (d), Cycle 23 (e), and Cycle 24 (f) and the linear speed of the CMEs along the X-ray peak flux during Cycle 23 and 24 (g), Cycle 23 (h), and Cycle 24 (i). The solar X-ray flare-associated regular CMEs are plotted as gray dots, while flare-associated halo CMEs are plotted as black dots. We binned the data with respect to the associated solar X-ray flares' peak flux to derive a log-linear correlation. Each bin, ordered from the lowest to highest flare peak flux, contains 94, 325, 744, 651, 372, and 274 samples for Cycle 23 and 24; 65, 217, 425, 429, 265, and 192 for Cycle 23; and 29, 108, 319, 222, 107, and 82 for Cycle 24. The median values of the bins are presented as blue dots. The error bars indicate the interquartile range (IQR) of the samples in the respective bins. The correlation derived from \cite{aarnio2011solar}, which includes LASCO CMEs from 1996 to 2006, is presented as a green solid line in (a), (b), and (c), whereas the best fit derived from the blue dots is presented as a blue solid line. The shaded region around each best-fit line represents the 1$\sigma$ confidence interval of the regression model, reflecting the uncertainty in the slope and intercept estimates.}
\label{fig:general}
\end{figure}

\begin{deluxetable*}{lclc}[!htbp]
\tablecaption{Correlation Equations Between X-ray Flare Peak Flux and CME Properties with Halo CMEs \label{tab:cme_corr}}
\tablehead{
\colhead{Solar Cycle} & \colhead{Derived Correlation Equations} & \colhead{$r_1$} & \colhead{$r_2$}
}
\startdata
23+24 & $\log M_{\rm CME} = (0.576\pm0.041)\, \log F_{\rm Xray} + (17.998\pm0.232)$ & 0.383 & 0.990 \\
23    & $\log M_{\rm CME} = (0.576\pm0.018)\, \log F_{\rm Xray} + (18.017\pm0.107)$ & 0.372 & 0.997 \\
24    & $\log M_{\rm CME} = (0.623\pm0.089)\, \log F_{\rm Xray} + (18.239\pm0.499)$ & 0.397 & 0.959 \\
23+24 & $\log E_{\rm CME} = (0.883\pm0.069)\, \log F_{\rm Xray} + (34.616\pm0.393)$ & 0.432 & 0.987 \\
23    & $\log E_{\rm CME} = (0.855\pm0.055)\, \log F_{\rm Xray} + (34.554\pm0.313)$ & 0.415 & 0.991 \\
24    & $\log E_{\rm CME} = (0.920\pm0.107)\, \log F_{\rm Xray} + (34.667\pm0.612)$ & 0.456 & 0.975 \\
23+24 & $\log v_{\rm CME} = (0.165\pm0.022)\, \log F_{\rm Xray} + (3.536\pm0.126)$   & 0.414 & 0.968 \\
23    & $\log v_{\rm CME} = (0.148\pm0.023)\, \log F_{\rm Xray} + (3.479\pm0.136)$   & 0.404 & 0.948 \\
24    & $\log v_{\rm CME} = (0.180\pm0.011)\, \log F_{\rm Xray} + (3.561\pm0.067)$   & 0.422 & 0.994 \\
\enddata
\tablecomments{The table shows regression results for CME mass ($M_{\rm CME}$), kinetic energy ($E_{\rm CME}$), and speed ($v_{\rm CME}$) as functions of soft X-ray flux ($F_{\rm Xray}$) presented in Figure 5. $r_1$ and $r_2$ are the Pearson coefficients for the full dataset and for the binned medians in Figure 5, respectively.}
\end{deluxetable*}

\begin{figure}[H]
\centering
\includegraphics[width=180mm]{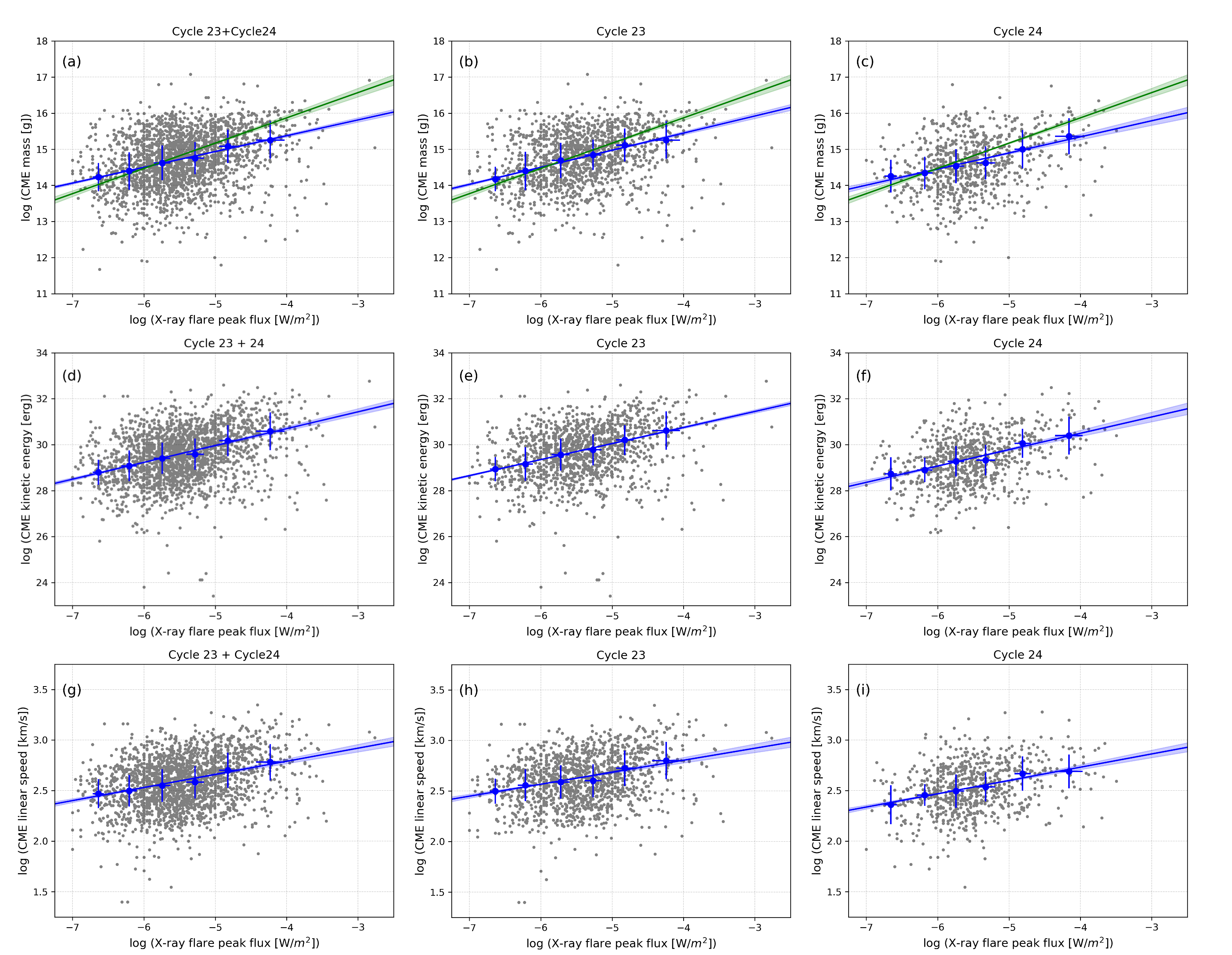}
\caption{The mass of the flare-associated LASCO CMEs without halo CMEs plotted along the X-ray peak flux of associated GOES X-ray flares that occurred during Solar Cycle 23 and 24 (a), Cycle 23 (b), and Cycle 24 (c), the kinetic energy of the CMEs along the X-ray peak flux during Cycle 23 and 24 (d), Cycle 23 (e), and Cycle 24 (f), and the linear speed of the CMEs along the X-ray peak flux during Cycle 23 and 24 (g), Cycle 23 (h), and Cycle 24 (i). The green solid line in (a), (b), and (c) shows the correlation derived from \cite{aarnio2011solar}. Each bin, ordered from the lowest to highest flare flux, contains 92, 311, 698, 592, 316, and 156 samples for Cycle 23 and 24; 63, 208, 397, 389, 228, and 111 for Cycle 23; and 29, 103, 301, 203, 88, and 45 for Cycle 24. The median values of the bins are presented as blue dots. The error bars indicate the interquartile range (IQR) of the samples in the respective bins.
\label{fig:general}}
\end{figure}

\begin{deluxetable*}{lclc}[!htbp]
\tablecaption{Correlation Equations Between X-ray Flare Peak Flux and CME Properties without Halo CMEs \label{tab:cme_corr}}
\tablehead{
\colhead{Solar Cycle} & \colhead{Derived Correlation Equations} & \colhead{$r_1$} & \colhead{$r_2$}
}
\startdata
23+24        & $\log M_{\rm CME} = (0.437\pm0.022)\, \log F_{\rm Xray} + (17.124\pm0.128)$  &0.293      & 0.995\\ 
23           & $\log M_{\rm CME} = (0.473\pm0.025)\, \log F_{\rm Xray} + (17.341\pm0.146)$  &0.290      & 0.993\\
24           & $\log M_{\rm CME} = (0.446\pm0.048)\, \log F_{\rm Xray} + (17.128\pm0.270)$  &0.287      & 0.979\\  
23+24        & $\log E_{\rm CME} = (0.733\pm0.050)\, \log F_{\rm Xray} + (33.631\pm0.286)$  &0.328      & 0.991\\  
23           & $\log E_{\rm CME} = (0.698\pm0.027)\, \log F_{\rm Xray} + (33.541\pm0.157)$  &0.313      & 0.997\\ 
24           & $\log E_{\rm CME} = (0.711\pm0.079)\, \log F_{\rm Xray} + (33.343\pm0.450)$  &0.343      & 0.979\\ 
23+24        & $\log v_{\rm CME} = (0.129\pm0.013)\, \log F_{\rm Xray} + (3.308\pm0.078)$   &0.297      & 0.979\\  
23           & $\log v_{\rm CME} = (0.118\pm0.015)\, \log F_{\rm Xray} + (3.278\pm0.088)$   &0.288      & 0.967\\ 
24           & $\log v_{\rm CME} = (0.131\pm0.014)\, \log F_{\rm Xray} + (3.257\pm0.079)$   &0.299      & 0.980\\ 
\enddata
\tablecomments{The equations and the Pearson coefficients of correlations presented in Figure 6.}
\end{deluxetable*}

\begin{figure}[H]
\centering
\includegraphics[width=180mm]{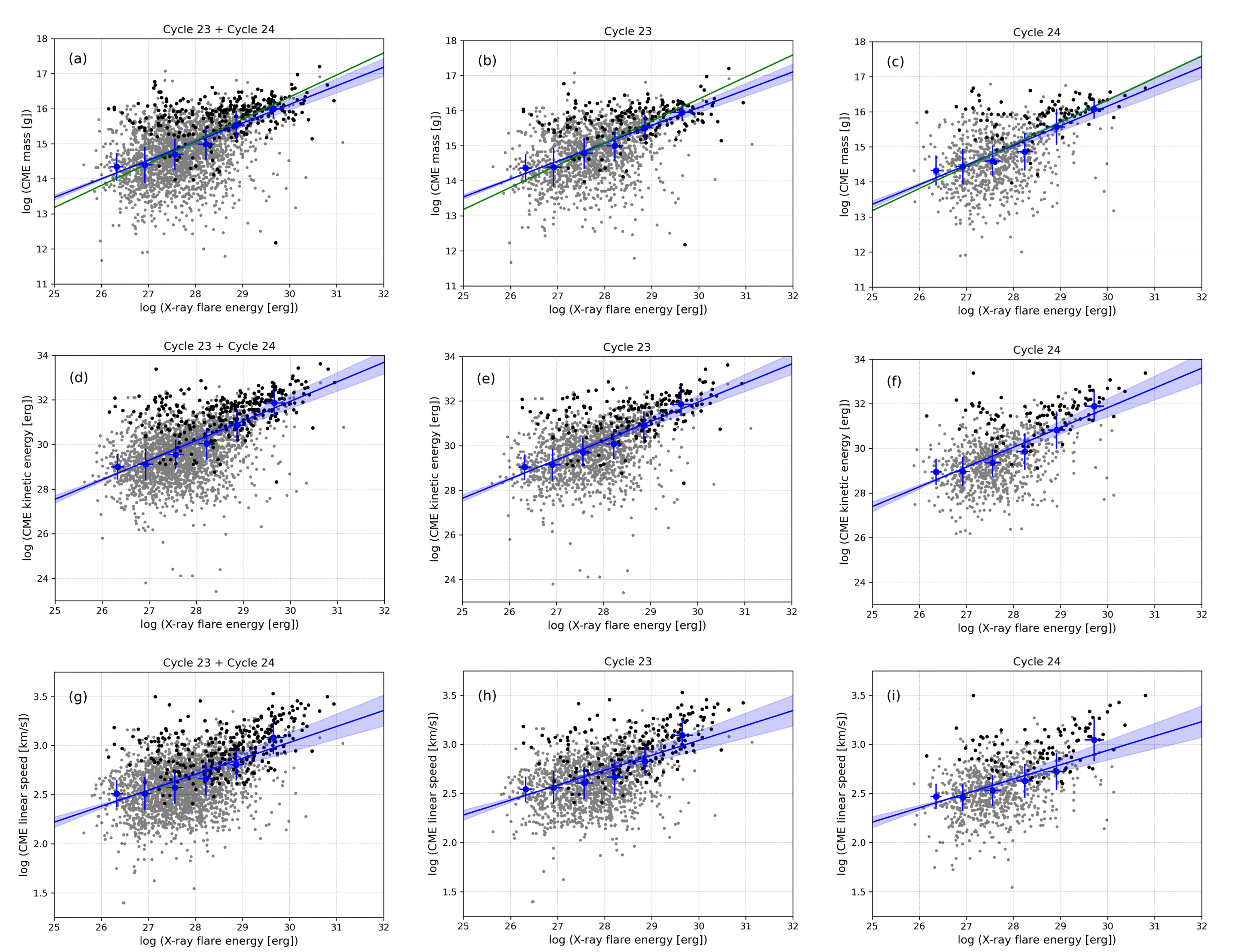}
\caption{The mass of the paired LASCO CMEs with halo CMEs plotted along the GOES X-ray flare energy of associated flares that occurred during Solar Cycle 23 and 24 (a), Cycle 23 (b), and Cycle 24 (c), the kinetic energy of the paired CMEs along the X-ray flare energy during Cycle 23 and 24 (d), Cycle 23 (e), and Cycle 24 (f), and the linear speed of the CMEs along the X-ray flare energy during Cycle 23 and 24 (g), Cycle 23 (h), and Cycle 24 (i). The green solid line in (a), (b), and (c) shows the correlation derived from \cite{aarnio2012mass}. Each bin, ordered from the lowest to highest flare energy, contains 117, 508, 796, 620, 294, and 125 samples for Cycle 23 and 24; 85, 312, 458, 440, 209, and 89 for Cycle 23; and 32, 196, 338, 180, 85, and 36 for Cycle 24. The median values of the bins are presented as blue dots. The error bars indicate the interquartile range (IQR) of the samples in the respective bins.
\label{fig:general}}
\end{figure}

\begin{deluxetable*}{lclc}[!htbp]
\tablecaption{Correlation Equations Between X-ray Flare Energy and CME Properties with Halo CMEs \label{tab:cme_corr}}
\tablehead{
\colhead{Solar Cycle} & \colhead{Derived Correlation Equations} & \colhead{$r_1$} & \colhead{$r_2$}
}
\startdata
23+24        & $\log M_{\rm CME} = (0.530\pm0.046)\, \log E_{\rm Xray} + (0.194\pm0.060)$  &0.431      & 0.979\\  
23           & $\log M_{\rm CME} = (0.509\pm0.041)\, \log E_{\rm Xray} + (0.800\pm0.053)$  &0.433      & 0.982\\  
24           & $\log M_{\rm CME} = (0.559\pm0.060)\, \log E_{\rm Xray} - (0.631\mp0.081)$  &0.423      & 0.964\\  
23+24        & $\log E_{\rm CME} = (0.878\pm0.097)\, \log E_{\rm Xray} + (5.574\pm0.118)$  &0.473      & 0.974\\ 
23           & $\log E_{\rm CME} = (0.860\pm0.089)\, \log E_{\rm Xray} + (6.121\pm0.107)$  &0.466      & 0.977\\  
24           & $\log E_{\rm CME} = (0.886\pm0.125)\, \log E_{\rm Xray} + (5.229\pm0.150)$  &0.478      & 0.963\\  
23+24        & $\log v_{\rm CME} = (0.162\pm0.030)\, \log E_{\rm Xray} - (1.843\mp0.035)$  &0.436      & 0.940\\  
23           & $\log v_{\rm CME} = (0.151\pm0.029)\, \log E_{\rm Xray} - (1.518\mp0.035)$  &0.434      & 0.929\\  
24           & $\log v_{\rm CME} = (0.146\pm0.030)\, \log E_{\rm Xray} - (1.445\mp0.035)$  &0.431      & 0.930\\  
\enddata
\tablecomments{The equations and the Pearson coefficients of correlations presented in Figure 7.}
\end{deluxetable*}

\begin{figure}[p]
\centering
\includegraphics[width=180mm]{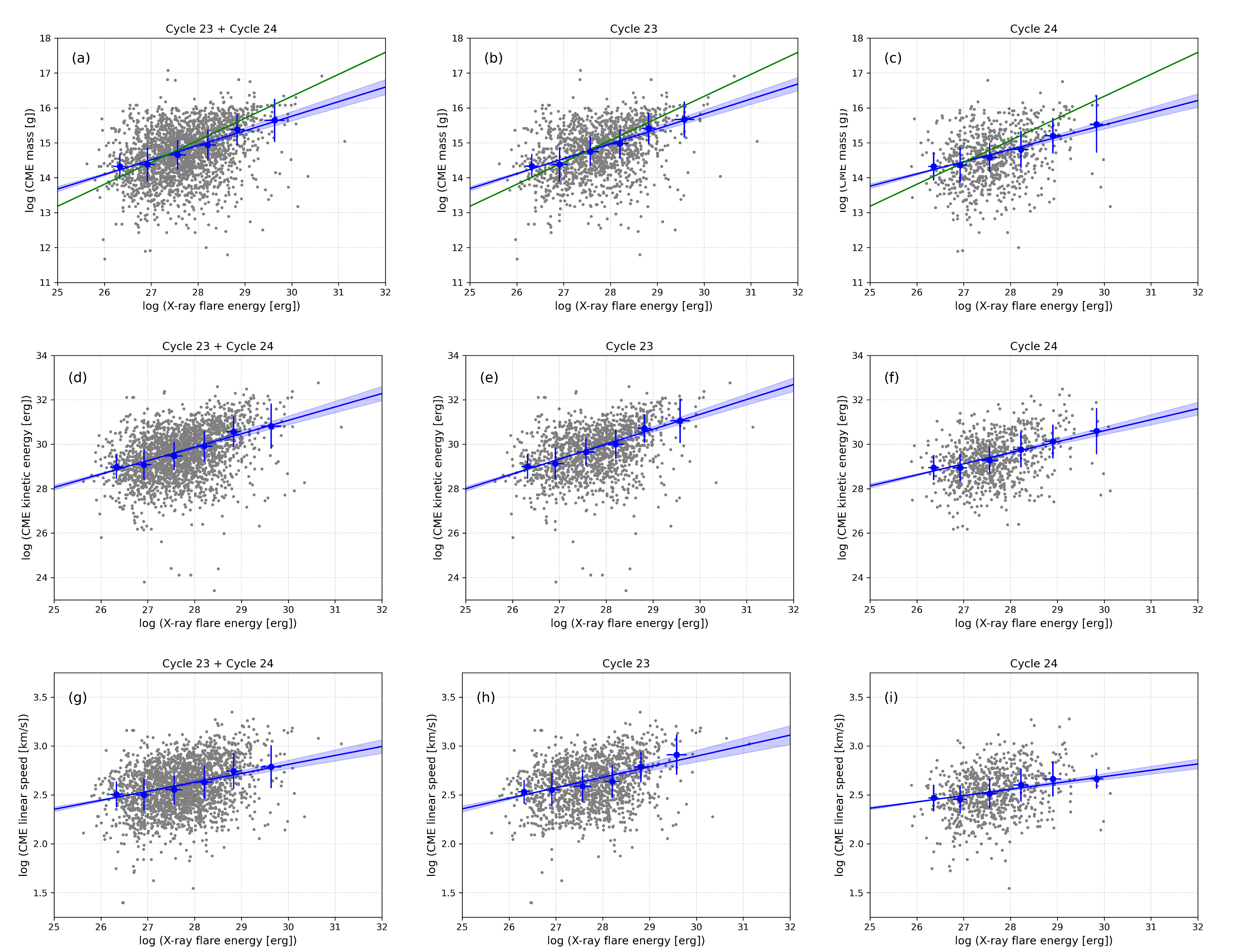}
\caption{The mass of the paired LASCO CMEs without halo CMEs plotted along the GOES X-ray energy of associated flares that occurred during Solar Cycle 23 and 24 (a), Cycle 23 (b), and Cycle 24 (c), the kinetic energy of the paired CMEs along the X-ray flare energy during Cycle 23 and 24 (d), Cycle 23 (e), and Cycle 24 (f), and the linear speed of the CMEs along the X-ray flare energy during Cycle 23 and 24 (g), Cycle 23 (h), and Cycle 24 (i). The green solid line in (a), (b), and (c) shows the correlation derived from \cite{aarnio2012mass}. Each bin, ordered from the lowest to highest flare energy, contains 113, 491, 741, 562, 216, and 42 samples for Cycle 23 and 24; 82, 302, 423, 399, 160, and 30 for Cycle 23; and 31, 189, 318, 163, 56, and 12 for Cycle 24. The median values of the bins are presented as blue dots. The error bars indicate the interquartile range (IQR) of the samples in the respective bins.
\label{fig:general}}
\end{figure}
\clearpage

\begin{deluxetable*}{lclc}[!htbp]
\tablecaption{Correlation Equations Between X-ray Flare Energy and CME Properties without Halo CMEs \label{tab:cme_corr}}
\tablehead{
\colhead{Solar Cycle} & \colhead{Derived Correlation Equations} & \colhead{$r_1$} & \colhead{$r_2$}
}
\startdata
23+24        & $\log M_{\rm CME} = (0.417\pm0.041)\, \log E_{\rm Xray} + (3.237\pm0.045)$  &0.337      & 0.985\\  
23           & $\log M_{\rm CME} = (0.428\pm0.036)\, \log E_{\rm Xray} + (2.982\pm0.040)$  &0.353      & 0.987\\  
24           & $\log M_{\rm CME} = (0.350\pm0.037)\, \log E_{\rm Xray} + (4.991\pm0.041)$  &0.294      & 0.986\\  
23+24        & $\log E_{\rm CME} = (0.604\pm0.061)\, \log E_{\rm Xray} + (12.942\pm0.067)$  &0.362      & 0.984\\  
23           & $\log E_{\rm CME} = (0.670\pm0.059)\, \log E_{\rm Xray} + (11.227\pm0.063)$  &0.364      & 0.988\\  
24           & $\log E_{\rm CME} = (0.496\pm0.055)\, \log E_{\rm Xray} + (15.708\pm0.062)$  &0.344      & 0.986\\  
23+24        & $\log v_{\rm CME} = (0.091\pm0.013)\, \log E_{\rm Xray} + (0.058\pm0.015)$  &0.309      & 0.970\\  
23           & $\log v_{\rm CME} = (0.107\pm0.013)\, \log E_{\rm Xray} - (0.338\mp0.020)$  &0.309      & 0.952\\  
24           & $\log v_{\rm CME} = (0.064\pm0.009)\, \log E_{\rm Xray} + (0.744\pm0.012)$  &0.291      & 0.951\\  
\enddata
\tablecomments{The equations and the Pearson coefficients of correlations presented in Figure 8.}
\end{deluxetable*}
\clearpage

\subsection{Correlation between FUV Peak Flux and the Mass of X-ray Flare-associated CMEs}

During the operation time of SDO (2010-present), we found 862 X-ray flare-CME pairs, including halo CMEs. In the one-minute averaged FUV light curve, we measure the peak FUV flux value within the event start and the end time of the corresponding X-ray flare event time. Among these pairs, a total of 743 CME-associated X-ray flares were recorded in the AIA1600 dataset. The loss of solar X-ray flares in the FUV dataset may have been a result of the operation schedule of SDO or a weak FUV flux during the solar X-ray flare event time. To predict the FUV peak flux of X-ray flares during Solar Cycle 23, we first investigated the relationships between the X-ray flare peak flux and the FUV flare peak flux and the X-ray flare energy and the FUV flare energy of simultaneously detected CME-associated X-ray flares during Solar Cycle 24, see Figure 9. The Pearson coefficient of the X-ray to FUV peak flux correlation is 0.747, and for the binned data, the coefficient is 0.974. The derived correlation follows

\begin{equation}
log F_{\rm FUV} = (0.512 \pm 0.065) log F_{\rm Xray} - (3.046 \mp 0.379)
\end{equation}

where $F_{\rm FUV}$ and $F_{\rm Xray}$ indicate the FUV and X-ray flare peak flux, respectively. The Pearson coefficient of the X-ray to FUV flare energy correlation is 0.847, and 0.997 for the binned data. The derived correlation for the flare energy follows

\begin{equation}
log E_{\rm FUV} = (0.669 \pm 0.026) log E_{\rm Xray} + (8.694 \pm 0.029)
\end{equation}

where $E_{\rm FUV}$ and $E_{\rm Xray}$ indicate the FUV and X-ray flare energy.

\begin{figure}[ht!]
\centering
\includegraphics[width=180mm]{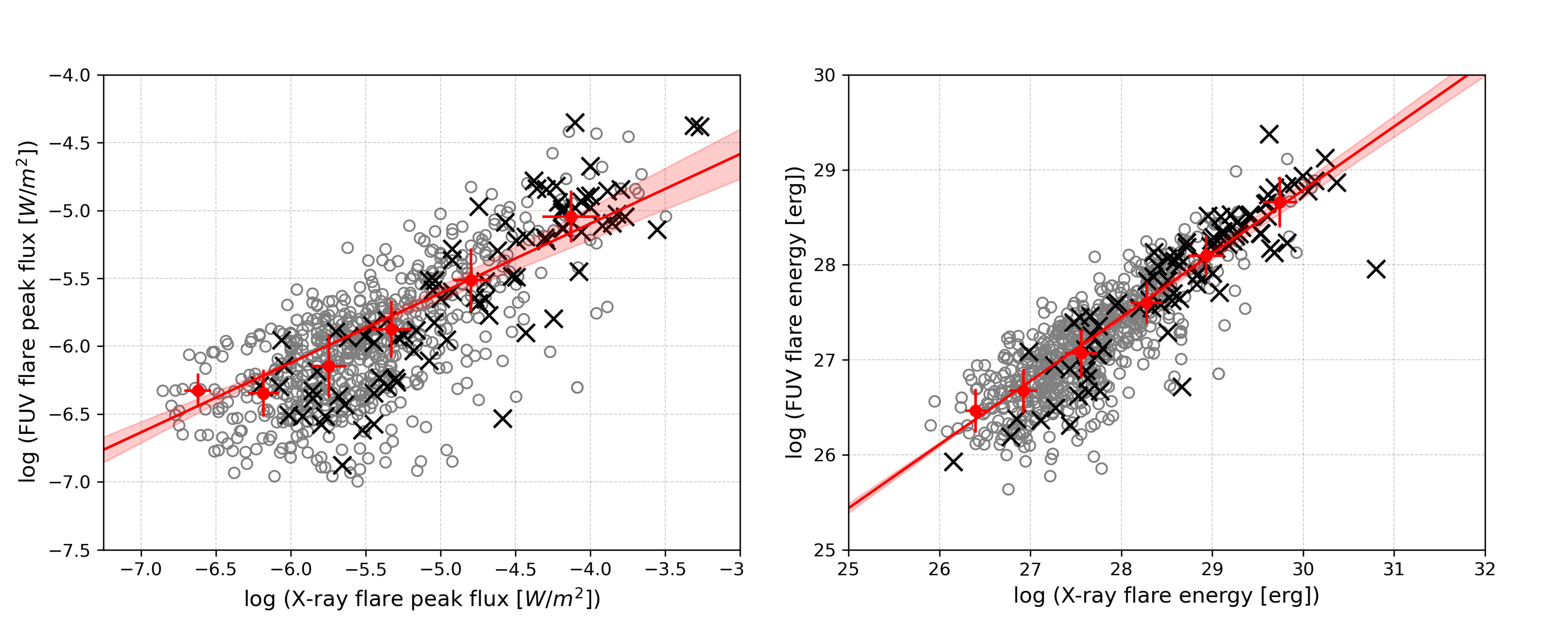}
\caption{The FUV flare peak flux [$W/m^{2}$] plotted against the X-ray flare peak flux [$W/m^{2}$] (left) and the FUV flare energy [erg] along the X-ray flare energy [erg] (right) during the simultaneous operation time of SDO and GOES (May/2010-Dec/2019). CME-associated X-ray flares recorded in the SDO AIA1600 instrument are plotted as open gray circles, and halo CME-associated flares are plotted as black cross markers. We binned the data with respect to the X-ray flare peak flux and derived a linear correlation. Each bin, ordered from the lowest to highest flare peak flux and energy, contains 20, 84, 269, 190, 100, and 80 samples for FUV-X-ray flux correlation; 23, 164, 276, 166, 80, and 34 for FUV-X-ray energy correlation. The median values of the bins are presented as red dots. The error bars on the bins show the interquartile range (IQR) of the samples in the corresponding bins. The red line is the linear correlation derived by the bins, whereas the shaded region around each best-fit line represents the 1$\sigma$ confidence interval of the regression model, which reflects the uncertainty in the slope and intercept estimates.
\label{fig:general}}
\end{figure}

These correlations can be applied to the CME-associated flares during Solar Cycle 23 and CME-associated flares that were not detected by SDO to convert the X-ray flare peak flux to the FUV flare peak flux and the X-ray flare energy to the FUV flare energy. Since excluding halo CMEs causes a considerable difference in flare-CME correlations, we assume that the halo CMEs are a physically different class of CME with different properties as \cite{kwon2015halo} suggested. By including FUV-X-ray correlation to the CME-associated X-ray flares, we find the FUV flare peak flux and the FUV flare energy of 2,460 CME-associated flares.

We found linear correlations between log FUV flare peak flux and log CME mass, kinetic energy, and linear speed, as well as between log FUV flare energy and log CME mass, kinetic energy, and linear speed for Solar Cycle 23 and Cycle 24, Cycle 23, and Cycle 24 (see Figures 10 and 11). The derived correlation equations and the Pearson coefficients of these correlations are presented in Tables 5 and 6. As FUV flare peak flux and energy distributions are slightly different from X-ray flare energy distributions, we used a 0.4 dex binning range for the FUV flare peak flux correlation and a 0.5 dex binning range for the flare energy correlation.

\begin{figure}[H]
\centering
\includegraphics[width=180mm]{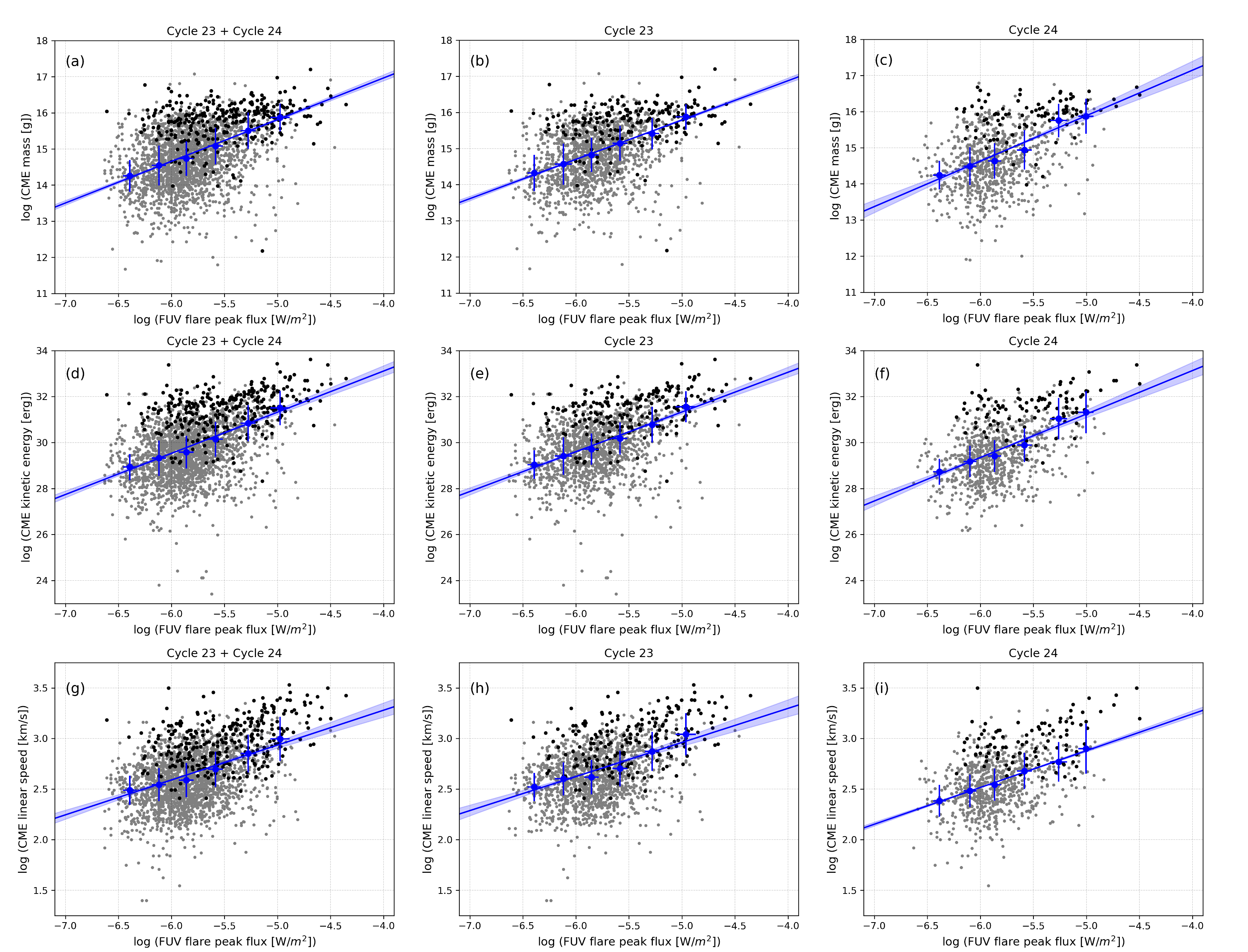}
\caption{The mass of X-ray flare-associated CMEs plotted along the FUV flare peak flux during Solar Cycle 23 and Cycle 24 (a), Cycle 23 (b), and Cycle 24 (c), the kinetic energy of paired CMEs along the FUV flare peak flux during Solar Cycle 23 and Cycle 24 (d), Cycle 23 (e), and Cycle 24 (f), and the linear speed of the CMEs along the FUV flare peak flux during Cycle 23 and 24 (g), Cycle 23 (h), and Cycle 24 (i). The FUV flare peak flux of CME-associated flares during Solar Cycle 23 is calculated by equating the FUV flare peak flux and the X-ray flare peak flux correlation derived from Figure 9. We binned the data with respect to the FUV flare peak flux to derive a linear correlation. Each bin, ordered from the lowest to highest flare peak flux, contains 157, 607, 842, 533, 219, and 102 samples for Cycle 23 and 24; 108, 364, 519, 380, 150, and 73 for Cycle 23; and 50, 243, 323, 153, 69, and 29 for Cycle 24. The median values of the bins are presented as blue dots. The error bars indicate the interquartile range (IQR) of the samples in the respective bins. The blue line is the linear correlation derived by the bins, whereas the uncertainty of the best fit is presented as a shaded region around the best fit. 
\label{fig:general}}
\end{figure}

\begin{deluxetable*}{lclc}[!htbp]
\tablecaption{Correlation Equations Between FUV Flare Peak Flux and CME Properties \label{tab:cme_corr}}
\tablehead{
\colhead{Solar Cycle} & \colhead{Derived Correlation Equations} & \colhead{$r_1$} & \colhead{$r_2$}
}
\startdata
23+24        & $\log M_{\rm CME} = (1.153\pm0.048)\, \log F_{\rm FUV} + (21.580\pm0.278)$  &0.383      & 0.996\\  
23           & $\log M_{\rm CME} = (1.088\pm0.043)\, \log F_{\rm FUV} + (21.231\pm0.245)$  &0.372      & 0.997\\  
24           & $\log M_{\rm CME} = (1.258\pm0.138)\, \log F_{\rm FUV} + (22.181\pm0.800)$  &0.397      & 0.974\\  
23+24        & $\log E_{\rm CME} = (1.790\pm0.123)\, \log F_{\rm FUV} + (40.273\pm0.717)$  &0.432      & 0.991\\  
23           & $\log E_{\rm CME} = (1.729\pm0.123)\, \log F_{\rm FUV} + (39.977\pm0.708)$  &0.415      & 0.990\\  
24           & $\log E_{\rm CME} = (1.891\pm0.187)\, \log F_{\rm FUV} + (40.697\pm1.100)$  &0.456      & 0.982\\  
23+24        & $\log v_{\rm CME} = (0.344\pm0.038)\, \log F_{\rm FUV} + (4.659\pm0.223)$  &0.414      & 0.980\\  
23           & $\log v_{\rm CME} = (0.336\pm0.045)\, \log F_{\rm FUV} + (4.645\pm0.266)$  &0.404      & 0.968\\  
24           & $\log v_{\rm CME} = (0.363\pm0.016)\, \log F_{\rm FUV} + (4.696\pm0.094)$  &0.422      & 0.996\\  
\enddata
\tablecomments{The equations and the Pearson coefficients of correlations presented in Figure 10.}
\end{deluxetable*}

\begin{figure}[H]
\centering
\includegraphics[width=180mm]{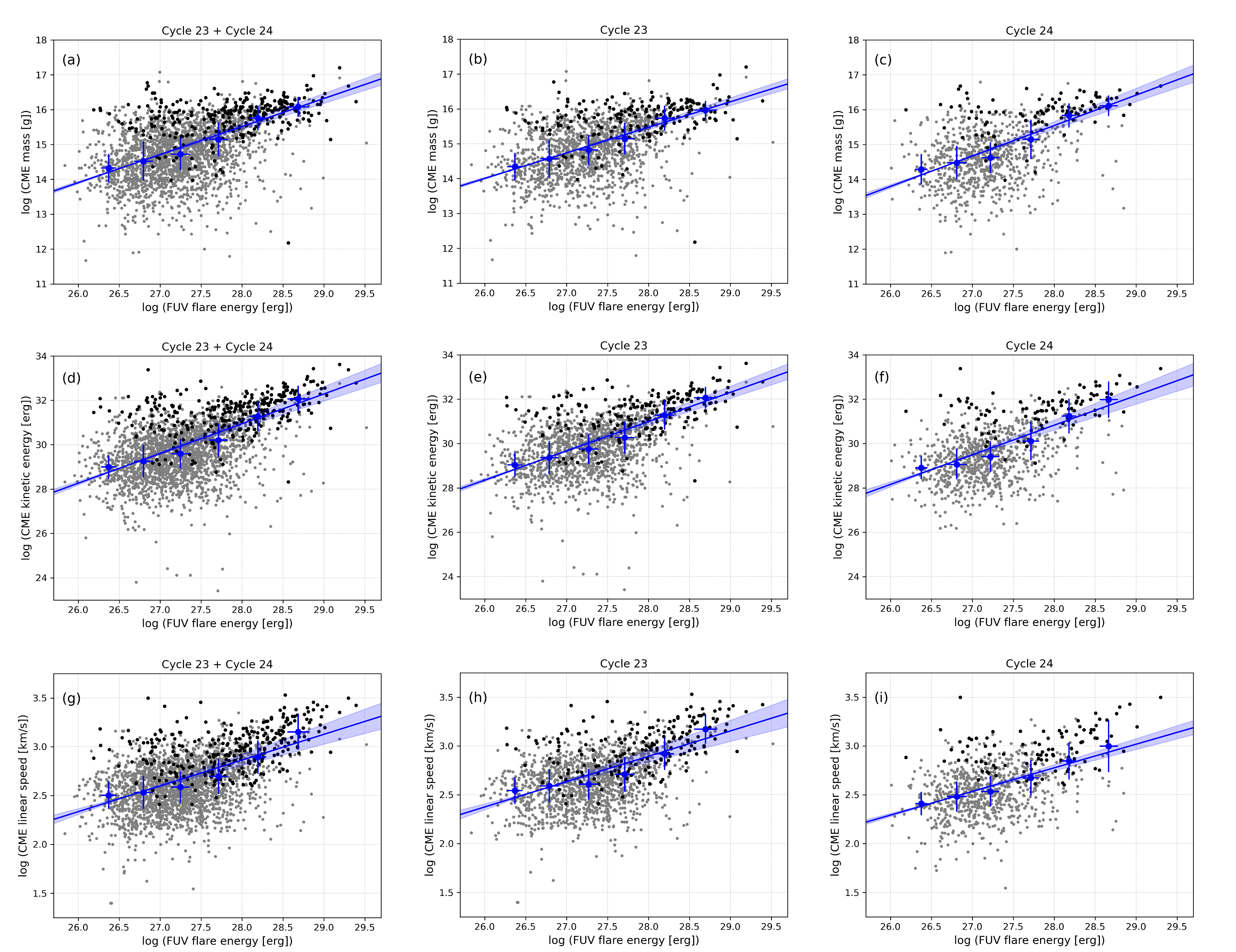}
\caption{The mass of X-ray flare-associated CMEs plotted along the FUV flare energy [erg] during Solar Cycle 23 and Cycle 24 (a), Cycle 23 (b), and Cycle 24 (c), the kinetic energy of paired CMEs along the FUV flare energy during Cycle 23 and Cycle 24 (d), Cycle 23 (e), and Cycle 24 (f), and the linear speed of the CMEs along the FUV flare energy during Cycle 23 and Cycle 24 (g), Cycle 23 (h), and Cycle 24 (i). Each bin, ordered from the lowest to highest flare energy, contains 167, 641, 816, 560, 199, and 76 samples for Cycle 23 and 24; 123, 377, 500, 402, 139, and 52 for Cycle 23; and 44, 264, 316, 158, 60, and 24 for Cycle 24. The median values of the bins are presented as blue dots. The error bars indicate the interquartile range (IQR) of the samples in the respective bins.
\label{fig:general}}
\end{figure}

\begin{deluxetable*}{lclc}[!htbp]
\tablecaption{Correlation Equations Between FUV Flare Energy and CME Properties \label{tab:cme_corr}}
\tablehead{
\colhead{Solar Cycle} & \colhead{Derived Correlation Equations} & \colhead{$r_1$} & \colhead{$r_2$}
}
\startdata
23+24        & $\log M_{\rm CME} = (0.804\pm0.059)\, \log E_{\rm FUV} - (7.020\mp0.052)$  &0.431      & 0.985\\  
23           & $\log M_{\rm CME} = (0.732\pm0.046)\, \log E_{\rm FUV} - (5.048\mp0.042)$  &0.433      & 0.990\\  
24           & $\log M_{\rm CME} = (0.873\pm0.080)\, \log E_{\rm FUV} - (8.907\mp0.069)$  &0.423      & 0.978\\  
23+24        & $\log E_{\rm CME} = (1.341\pm0.138)\, \log E_{\rm FUV} - (6.623\mp0.119)$  &0.473      & 0.977\\  
23           & $\log E_{\rm CME} = (1.318\pm0.114)\, \log E_{\rm FUV} - (5.942\mp0.100)$  &0.466      & 0.982\\  
24           & $\log E_{\rm CME} = (1.336\pm0.167)\, \log E_{\rm FUV} - (6.584\mp0.140)$  &0.478      & 0.973\\  
23+24        & $\log v_{\rm CME} = (0.264\pm0.044)\, \log E_{\rm FUV} - (4.533\mp0.035)$  &0.436      & 0.950\\  
23           & $\log v_{\rm CME} = (0.259\pm0.046)\, \log E_{\rm FUV} - (4.376\mp0.038)$  &0.434      & 0.937\\  
24           & $\log v_{\rm CME} = (0.242\pm0.024)\, \log E_{\rm FUV} - (4.003\mp0.019)$  &0.431      & 0.985\\
\enddata
\tablecomments{The equations and the Pearson coefficients of correlations presented in Figure 11.}
\end{deluxetable*}

\clearpage

\section{Conclusion} \label{sec:result}
Since stellar CMEs can affect the stability of planetary atmospheres, constraining the properties of stellar CMEs provides additional information to better assess the potential habitability of exoplanets. Since observations of stellar CMEs have not yet produced conclusive detections, we used previously reported solar X-ray flare-CME association to establish constraints on stellar CMEs in FUV, where the flares are actively observed. We performed a series of correlation analyses that include X-ray flare peak flux-CME mass, flare peak flux-CME kinetic energy,  flare peak flux-CME linear speed, flare energy-CME mass, flare energy-CME kinetic energy, and flare energy-CME linear speed. In the process, we compiled X-ray flares and CMEs available for the last 23 years, from 1996 to 2019. Among these, we found 2,460 X-ray flare-associated CMEs by applying temporal and spatial constraints from \cite{aarnio2011solar}. The log-linear correlation derived in this study agrees with the correlation reported by \cite{aarnio2011solar} and \cite{aarnio2012mass} when halo CMEs are included. When halo CMEs are excluded, the correlations diverge from those obtained when halo CMEs are included. This discrepancy is most evident at the high-flux and high-energy end (above $10^{-5} [W/m^2]$ for flare flux and $10^{28}[erg]$ for flare energy), where they predict systematically lower CME masses, kinetic energies, and linear speeds for a given flare peak flux or flare energy. This emphasizes the role of halo CMEs in predicting the mass of stellar CMEs with associated flare observation. We also found that the correlations of flare-CME pair properties explored in this study are not disturbed by the Solar Cycles, showing relatively consistent Pearson coefficients. This implies that estimating CME properties from flare observations appears insensitive to stellar cycle variations.

Using AIA1600/SDO FUV images, we converted the pixel values of the corresponding image to spectral irradiance data (1600 \AA). Then, the FUV peak flux during the solar X-ray flare event was extracted to be correlated with X-ray flare-associated CME mass, kinetic energy, and linear speed. Among 862 solar X-ray flare-CME pairs during the SDO operation period, we found 743 FUV flare peak flux of the CME-associated X-ray flares and derived their energy. To incorporate CME-associated X-ray flares during Solar Cycle 23, we found a correlation between the X-ray flare peak flux and FUV flare peak flux of CME-associated X-ray flares and a correlation of their respective energy as well. We applied these correlations to the CME-associated X-ray flares during Solar Cycle 23. In total, we used 2,460 flare-CME pairs from 1996 to 2019 and found a series of log-linear correlations. 

While candidate stellar CMEs have been observed, none have been confirmed for Sun-like stars with simultaneous measurements of both flare X-ray flux and CME mass. This gap is primarily due to observational limitations: no current observatories are designed specifically to detect stellar CMEs, particularly around solar analogs \citep{lloyd2025}. Coronagraphy, which enables CME mass measurements for the Sun, is not yet viable for distant stars due to contrast and spatial resolution challenges. Moreover, Sun-like stars are generally less magnetically active than M dwarfs, leading to weaker flares and potentially less frequent or less massive CMEs, making detections harder. Addressing this will require new instruments optimized for stellar space weather diagnostics.

\section*{Acknowledgments}
This material is based upon work performed as part of the CHAMPs (Consortium on Habitability and Atmospheres of M-dwarf Planets) team, supported by the National Aeronautics and Space Administration (NASA) under Grant Nos. 80NSSC21K0905 and 80NSSC23K1399 issued through the Interdisciplinary Consortia for Astrobiology Research (ICAR) program. The CME catalog used in this paper is generated and maintained at the CDAW Data Center by NASA and the Catholic University of America in Cooperation with the Naval Research Laboratory. SOHO is a project of international cooperation between ESA and NASA. We also thank the thoughtful questions and comments from the anonymous referee.

\bibliography{sample631}{}
\bibliographystyle{aasjournal}
\end{document}